\documentclass[aps,prd,preprint,superscriptaddress]{revtex4}

\usepackage{longtable}
\usepackage{graphicx}
\usepackage{color}
\usepackage{bbm}
\usepackage{subfigure}
\usepackage{amsmath,amssymb}

\newcommand{\e}{{\rm e}}
\newcommand{\irm}{{\rm i}}
\newcommand{\beq}{\begin{equation}}
\newcommand{\eeq}{\end{equation}}
\newcommand{\bdm}{\begin{displaymath}}
\newcommand{\edm}{\end{displaymath}}

\begin{document}

\title{Seismic topographic scattering in the context of GW detector site selection}

\author{M Coughlin}
\address{Carleton College, Northfield, MN 55057, USA}
\author{J Harms}
\address{California Institute of Technology, Pasadena, California 91125, USA}

\begin{abstract}
In this paper, we present a calculation of seismic scattering from irregular surface topography in the Born approximation. Based on US-wide topographic data, we investigate topographic scattering at specific sites to demonstrate its impact on Newtonian-noise estimation and subtraction for future gravitational-wave detectors. We find that topographic scattering at a comparatively flat site in Oregon would not pose any problems, whereas scattering at a second site in Montana leads to significant broadening of wave amplitudes in wavenumber space that would make Newtonian-noise subtraction very challenging. Therefore, it is shown that topographic scattering should be included as criterion in the site-selection process of future low-frequency gravitational-wave detectors.
\end{abstract}
\pacs{04.80.Nn,91.30.Fn,95.75.Wx}

\maketitle

\section{Introduction}
\label{sec:Intro}
Seismic disturbances are an important limitation to the sensitivity of ground-based gravitational-wave (GW) detectors. Sophisticated seismic isolation systems will be used in next-generation detectors like Advanced LIGO \cite{aLIG2011}, Advanced Virgo \cite{aVir2009} or LCGT \cite{KuEA2010} to push the sensitive band down to frequencies around 10\,Hz. The goal of future-generation detectors like the European Einstein Telescope \cite{HiEA2011,ET2011} will be to extend the detection band to even lower frequencies like 3\,Hz and below. The detectors need to be stabilized against occasional seismic disturbances, as for example those generated by earthquakes, but the ambient seismic field also generates a stationary background noise that cannot be fully avoided despite multi-stage passive and active isolation systems.

Seismic fields also generate perturbations of the gravity field that will contribute to the instrumental noise as Newtonian noise \cite{HuTh1998,BeEA1998,Cre2008,ThWi1999}. These perturbations are extremely weak, and they will only be significant noise contributions in future detectors beyond Advanced LIGO and Advanced Virgo \cite{BeEA2010}. Whereas it is possible to conceive further improvements to current seismic isolation designs, it is not possible to shield the motion of the suspended test masses from fluctuations of the gravity field. The best strategy is to search for detector sites that have a weak seismic spectrum within the relevant frequency band between 1\,Hz and 20\,Hz \cite{HaEA2010}. In addition, because the sources of the gravity perturbations are known, it should be possible to estimate Newtonian noise using data from seismic arrays and then subtract the estimate coherently from the GW data \cite{HaEA2009a,HaEA2009b}. Seismic fields are difficult to analyze at the surface due to heterogeneities of the near-surface layers \cite{HaOR2011} and abundance of seismic sources. Therefore, constructing future detectors underground would not only provide remote, low-noise environments for the detectors but also improve and simplify the Newtonian-noise estimation, potentially yielding smaller subtraction residuals. 

In this paper, we address a specific problem that is related to the Newtonian-noise estimation. The claimed advantage of underground sites is partially based on the assumption that the medium is more homogeneous at greater depths, and local seismic sources would be rare or weak. This is certainly the case if underground sites are chosen carefully, but another effect that causes heterogeneity in the seismic field is the scattering of seismic fields from an irregular surface topography. We will first outline the theory of seismic scattering in the Born approximation in section \ref{sec:topotheory} and explicitly evaluate scattering coefficients for a few examples. We then present topographical data for the US in section \ref{sec:topodata} and use a simple figure-of-merit to select interesting sites. A detailed scattering analysis is carried out for these sites in section \ref{sec:scattsite}. The paper concludes in section \ref{sec:Conclude}, where we discuss implications of our results for Newtonian-noise estimation and subtraction.

\section{Topographic scattering: theory}
\label{sec:topotheory} 
The theory of seismic-wave topographic scattering in the Born approximation was first developed for the 2D case in \cite{GiKn1960} and more specifically for sinusoidal surfaces in \cite{Abu1962}. Later, the 3D case was discussed in \cite{Hud1967}. The Born approximation is convenient for the purpose of this paper, but it should be noted that alternative approximations have been developed that hold for a wider class of topographies (see \cite{Ogi1987} for details). In this section, we will outline the basic ideas and fundamental equations. We will consider sinusoidal surfaces in some detail, applying a more efficient approach compared to \cite{Abu1962} by formulating the problem in terms of boundary conditions instead of effective source distributions at the surface. Scattering coefficients will then be plotted for the most common types of incident plane waves. 

The scenario that we consider is that of plane seismic waves propagating on or inside a homogeneous medium with arbitrary surface topography. If the surface was flat, then surface waves would propagate without conversion into other seismic modes. Seismic waves propagating through the medium and reflecting from a flat surface experience partial conversion between compressional and shear modes. In each case, we denote the total displacement field throughout the entire medium by ${\vec\xi}^{\,0}(\vec x,t)$. The superscript ``0'' indicates that we consider the displacement field assuming a flat surface. The waves emanating from a plane surface are called specular reflection. As we will show, an incident field is scattered from an irregular surface topography. In this case, we decompose the total displacement field $\vec \xi$ into its unperturbed component ${\vec\xi}^{\,0}$ and the scattered field ${\vec\xi}^{\,\rm s}$. We assume that further scattering of a scattered wave is negligible, i.~e.~the calculations are carried out in the Born approximation. 

The surface topography shall be described by a function $z=s(x,y)$, which is treated as a perturbation of a flat surface such that the mean value vanishes: $\langle s(x,y)\rangle = 0$. The traction at a free surface, which is the stress tensor projected onto the normal direction  $\vec n\equiv n_i\vec e_i$ of the surface, vanishes (see for example \cite{AkRi2009}):
\beq
\tau_{ij}n_i\equiv \tau_{nj}=0
\label{eq:freetraction}
\eeq
We desire to translate these boundary conditions defined on a rough surface to equivalent conditions on the (virtual) plane $z=0$. The new boundary conditions must involve non-vanishing surface traction depending on $s(x,y)$. To calculate the surface traction on the plane, we first rotate the traction of the free surface $s(x,y)$ around the two coordinate axes $x,\,y$ with rotation angles $\alpha,\,\beta$ obtained from 
\beq
\tan\alpha(x,y)=-\partial_ys(x,y),\quad\tan\beta(x,y)=\partial_xs(x,y)
\eeq
assuming that the rotation angles are small. Next, we translate the surface point along the $z$-axis from $z=s(x,y)$ to $z=0$. If the elevation is small, then we achieve this through a Taylor expansion of the stress tensor according to:
\beq
\tau_{ij}(x,y,z=s(x,y))\approx\tau_{ij}(x,y,0)+s(x,y)\partial_z\tau_{ij}(x,y,z)\Big|_{z=0}
\label{eq:stressexpand}
\eeq
Whether the last equation is a good approximation also depends on the length of seismic waves (and therefore frequency) since $\tau_{ij}$ is the stress tensor associated with some displacement field. The greater the seismic wavelength in the vertical direction (along the $z$-axis), the smaller $\partial_z\tau_{ij}$, resulting in a better approximation. Finally, analogous with the displacement field, the stress field will be divided into an unperturbed stress $\tau_{ij}^0$ in case of a flat surface and a stress tensor $\tau_{ij}^{\rm s}$ associated with the scattered field. Neglecting higher-order perturbations in $s(x,y)$ and combining equations (\ref{eq:freetraction}) and (\ref{eq:stressexpand}), we find
\begin{eqnarray}
\tau_{xz}^{\rm s} &= -s(x,y)\partial_z\tau_{xz}^0+(\partial_x s(x,y))\tau_{xx}^0+(\partial_y s(x,y))\tau_{xy}^0 \label{eq:scattx}\\
\tau_{yz}^{\rm s} &= -s(x,y)\partial_z\tau_{yz}^0+(\partial_x s(x,y))\tau_{xy}^0+(\partial_y s(x,y))\tau_{yy}^0 \label{eq:scatty}\\
\tau_{zz}^{\rm s} &= -s(x,y)\partial_z\tau_{zz}^0 \label{eq:scattz}
\end{eqnarray}
where all stress tensors being evaluated on the plane $z=0$. We used $\tau_{iz}^0(x,y,0)=0$ because the unperturbed stress field produces a vanishing traction normal to a free, flat surface. Linearizing the surface stress equations with respect to the surface profile $s(x,y)$ is part of the Born approximation.

We now have equations that allow us to calculate a stress field $\tau_{jz}^{\rm s}$ at $z=0$ in terms of a surface topography $s(x,y)$ and unperturbed stress $\tau_{ij}^0$. The two remaining problems are to calculate the unperturbed stress field $\tau_{ij}^0$ and its derivatives (at $z=0$) given an unperturbed seismic field ${\vec\xi}^{\,0}$ and then to calculate the scattered displacement field ${\vec\xi}^{\,\rm s}$ throughout the entire medium using boundary conditions that are derived from the effective surface loads $\tau_{iz}^{\rm s}$.
For this purpose, it is convenient to express the unperturbed displacement field in terms of potentials
\beq
{\vec\xi}^{\,0}(\vec x,t) = \nabla\phi(\vec x,t)+\nabla\times\vec\psi(\vec x,t)
\label{eq:potentials}
\eeq
as we can now easily associate the scalar potential $\phi$ with compressional modes and the vector potential $\vec\psi$ with shear modes ($\nabla\cdot\vec\psi=0$). For each of the potentials, the traction with respect to planes perpendicular to the $z$-axis reads
\begin{eqnarray}
\tau^{\rm cmp}_{iz} &= \lambda\Delta\phi\,\delta_{iz}+2\mu\partial_z\partial_i\phi
\label{eq:stresscompr}\\
\tau^{\rm shr}_{iz} &= \mu\left(2\partial_z(\nabla\times\vec\psi)-\Delta(\vec e_z\times\vec\psi)\right)_i
\label{eq:stressshear}
\end{eqnarray}
where the subscript ``$i$'' in the last equation denotes the component $i$ of the vector inside the brackets, $\lambda,\,\mu$ are the Lam\'e constants, and $\Delta$ is the Laplace operator. We will not give explicit expressions here for the potentials. They can be found in \cite{GiKn1960,AkRi2009}. We just want to point out that in all cases studied in this paper, i.~e.~scattering of incident compressional, shear and fundamental Rayleigh waves, the scalar and vector potentials both need to be included due to partial mode conversion at reflection from the flat surface (the Rayleigh field being a complex continuation of the bulk-wave reflection). 

The same equations (\ref{eq:stresscompr}) and (\ref{eq:stressshear}) together with equation (\ref{eq:potentials}) and $\nabla\cdot\vec\psi=0$ can be used to calculate the scattered waves. One starts with the surface load, equations (\ref{eq:scattx}) to (\ref{eq:scattz}), and solves for the amplitudes of the field potentials. In other words, the surface load forms a boundary condition for the scattered field. We can determine the potentials analytically only for very simple surface loads. In this paper, we calculate the amplitudes of plane waves contained in the scattered field as a function of the wave vector $\vec\kappa$ that characterizes a sinusoidal surface profile and for different types of incident plane waves. Then the scattering problem is analogous to reflection of light from optical gratings, and results can be easily interpreted in wavenumber space. If the $s(x,y)$ has a continuous spectrum, one needs to integrate over all possible horizontal wavenumbers of the scattered field, but because we consider load distributions $\tau_{iz}^{\rm s}(x,y)$ with discrete spatial spectrum and since we understand the surface load as boundary condition of the scattered field (as opposed to a source distribution), no integration is required to calculate the amplitudes of the scattered waves. 

As an example, let us consider an incident plane compressional wave. Its wave vector can be decomposed into a sum over the horizontal (along the plane $z=0$) and vertical wave vectors: $\vec k^{\,0}=\vec k^{\,0,\,\rm h}+\vec k^{\,0,\,\rm v}$. At reflection from a flat surface, the horizontal wave vector is preserved among all in and outgoing waves, but the vertical wave vector will have different values for the different wave modes. There is a simple sign flip in $\vec k^{\,0,\,\rm v}$ for the reflected compressional wave, but the shear wave generated at reflection has a vertical wavenumber $k_{\rm shr}^{\,0,\,\rm v}>k^{\,0,\,\rm v}$ because the shear wave speed is smaller than the compressional wave speed. So in this case, the complete unperturbed displacement field including the incident and reflected waves is characterized by a single horizontal wave vector and three vertical wave vectors. According to equations (\ref{eq:stresscompr}) and (\ref{eq:stressshear}), the same wave vectors also characterize the unperturbed stress field. Substituting a sinusoidal function $\exp(\pm\irm\vec\kappa\cdot\vec x)$ for the surface topography $s(x,y)$ in equations (\ref{eq:scattx}) to (\ref{eq:scattz}), we find that the surface load is characterized by a discrete spectrum with horizontal wave vectors $\vec k^{\,\rm s,\, h}_\pm=\vec k^{\,0,\,\rm h}\pm\vec\kappa$. It clearly follows that the scattered waves must have the same horizontal wave vectors. Therefore, in this simple example, the scattered field is composed of waves with four different wave vectors determined by the two $\vec k^{\,\rm s,\, h}_\pm$ and two different wave speeds $\alpha,\,\beta$ associated with compressional and shear waves. The scattered field can be represented in the form
\begin{eqnarray}
{\vec\xi}^{\,\rm s}(\vec x,t)=&\vec C(\vec k^{\,\rm s,\, h}_+)\e^{\irm(\vec k^{\,\rm s}_+(\alpha)\cdot\vec x-\omega t)}+\vec C(\vec k^{\,\rm s,\, h}_-)\e^{\irm(\vec k^{\,\rm s}_-(\alpha)\cdot\vec x-\omega t)}\nonumber\\
&+\vec S(\vec k^{\,\rm s,\, h}_+)\e^{\irm(\vec k^{\,\rm s}_+(\beta)\cdot\vec x-\omega t)}+\vec S(\vec k^{\,\rm s,\, h}_-)\e^{\irm(\vec k^{\,\rm s}_-(\beta)\cdot\vec x-\omega t)}
\end{eqnarray}
where the 3D wave vectors $\vec k^{\,\rm s}_\pm$ have vertical wavenumber 
\beq
k^{\,\rm s,\,v}_\pm(c)\equiv\sqrt{\frac{\omega^2}{c^2}-(k^{\,\rm s,\,h}_\pm)^2}
\eeq
The scattering coefficients $\vec C,\,\vec S$ describe the scattering of the incident wave into compressional and shear waves. Small-wavelength sinusoidal surfaces, $\kappa\gg k^{\,0,\,\rm h}$, scatter waves predominantly at large angles with respect to the specular reflection, whereas surfaces with $\kappa\ll k^{\,0,\,\rm h}$ scatter waves at small angles relative to the specular reflections. In general, the topographic spectrum will be continuous and extend to arbitrarily large wavenumbers $\kappa$. Therefore, the wavenumber spectrum of the scattered field will extend to arbitrarily large horizontal wavenumbers $k^{\,\rm s,\, h}$, and a bulk compressional or shear wave can scatter into a surface wave with imaginary vertical wavenumber and vice versa. Among the scattered surface waves, only the ones with a specific horizontal wavenumber correspond to the fundamental Rayleigh mode, which means that scattering creates surface waves that are not Rayleigh waves. These waves cannot propagate freely because fundamental Rayleigh waves are the only surface waves that can propagate freely on homogenous media. In the context of Newtonian noise estimation, scattered surface waves that have wavelengths much shorter than Rayleigh waves can pose a problem. If these waves have significant amplitude, very dense seismic arrays would be necessary to monitor the surface displacement. 

\begin{figure*}[t]
 \centering \subfigure{}
 \includegraphics[width=3in]{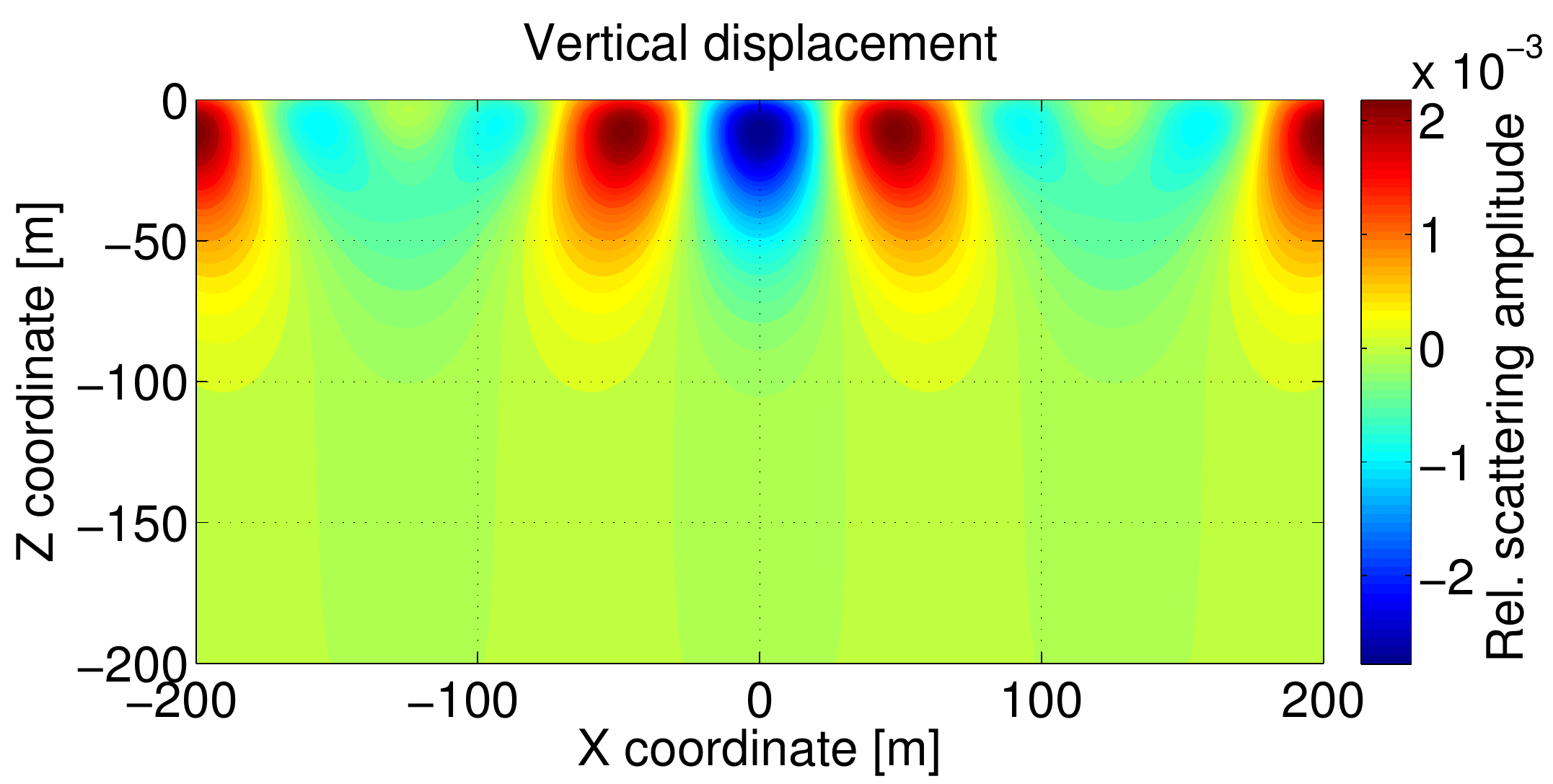}
 \centering \subfigure{}
 \includegraphics[width=3in]{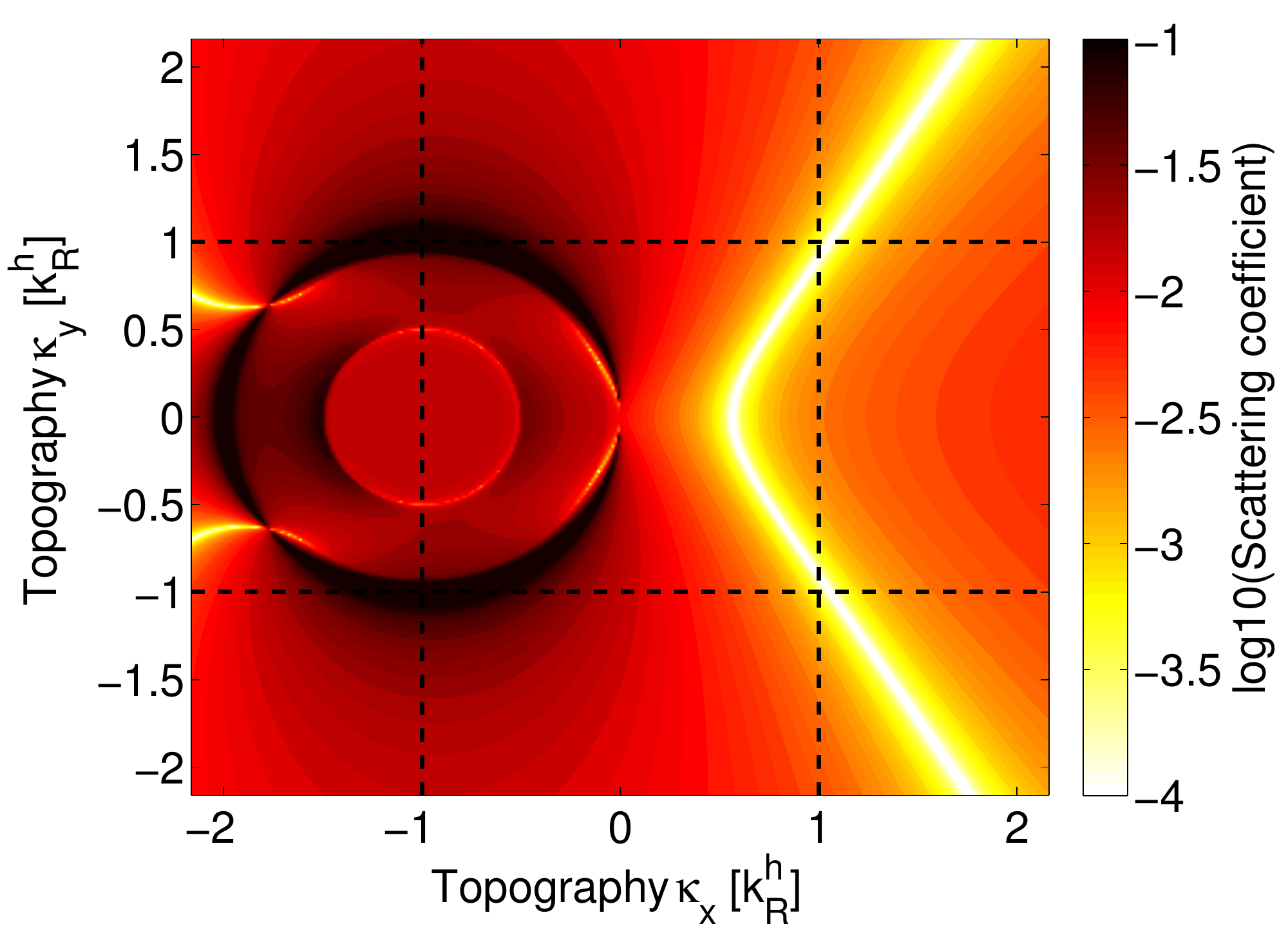}
 \caption{The plot to the left shows an example of a scattered displacement field in the $x,z$-plane. The incident wave is a Rayleigh wave with horizontal wavelength $\lambda^{\rm h}_{\rm inc}\approx 500\,$m propagating along the x-direction. It is scattered from a sinusoidal surface that has a wavelength $\lambda_{\rm s}\equiv 2\pi/\kappa=100\,$m. The incident wave has vertical amplitude 1\,m at $z=0$ and the surface topography has amplitude 1\,m. In the Born approximation, the amplitude of the scattered wave scales linearly with the amplitude of the surface profile and linearly with the amplitude of the incident wave. One can see that the evanescent field is not of Rayleigh type because its phase is a function of $z$. The plot to the right shows the absolute value of the vertical displacement of the scattered field at the surface normalized by the amplitudes of the topography and incident wave. The topographic wave vector is expressed in units of the horizontal wavenumber of the incident Rayleigh wave. A resonance effect between scattered shear and compressional waves causes the scattering amplitude to go to infinity for certain $\vec \kappa$, which lie on the dark colored circle. The Born approximation breaks down at these wave vectors, which are known as the Rayleigh poles. For this reason, it is not possible to calculate a total integrated scatter for harmonic topographies in the Born approximation. The problem seems to be analogous to the emergence of secular terms in perturbation theory that is often solved by including dispersion terms in the solution.}
\label{fig:scattR}
\end{figure*}

\begin{figure*}[t]
 \centering \subfigure{}
 \includegraphics[width=3in]{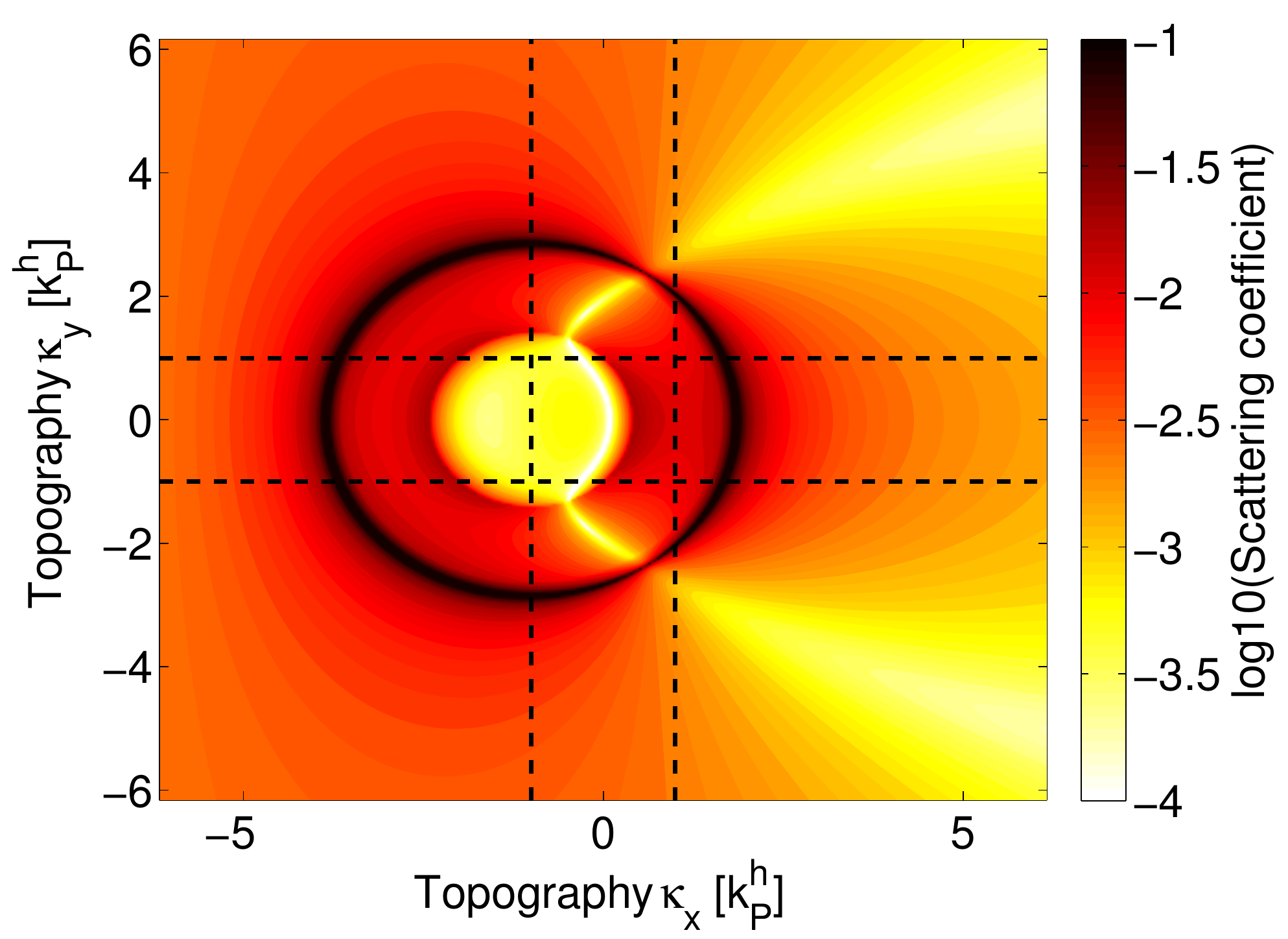}
 \centering \subfigure{}
 \includegraphics[width=3in]{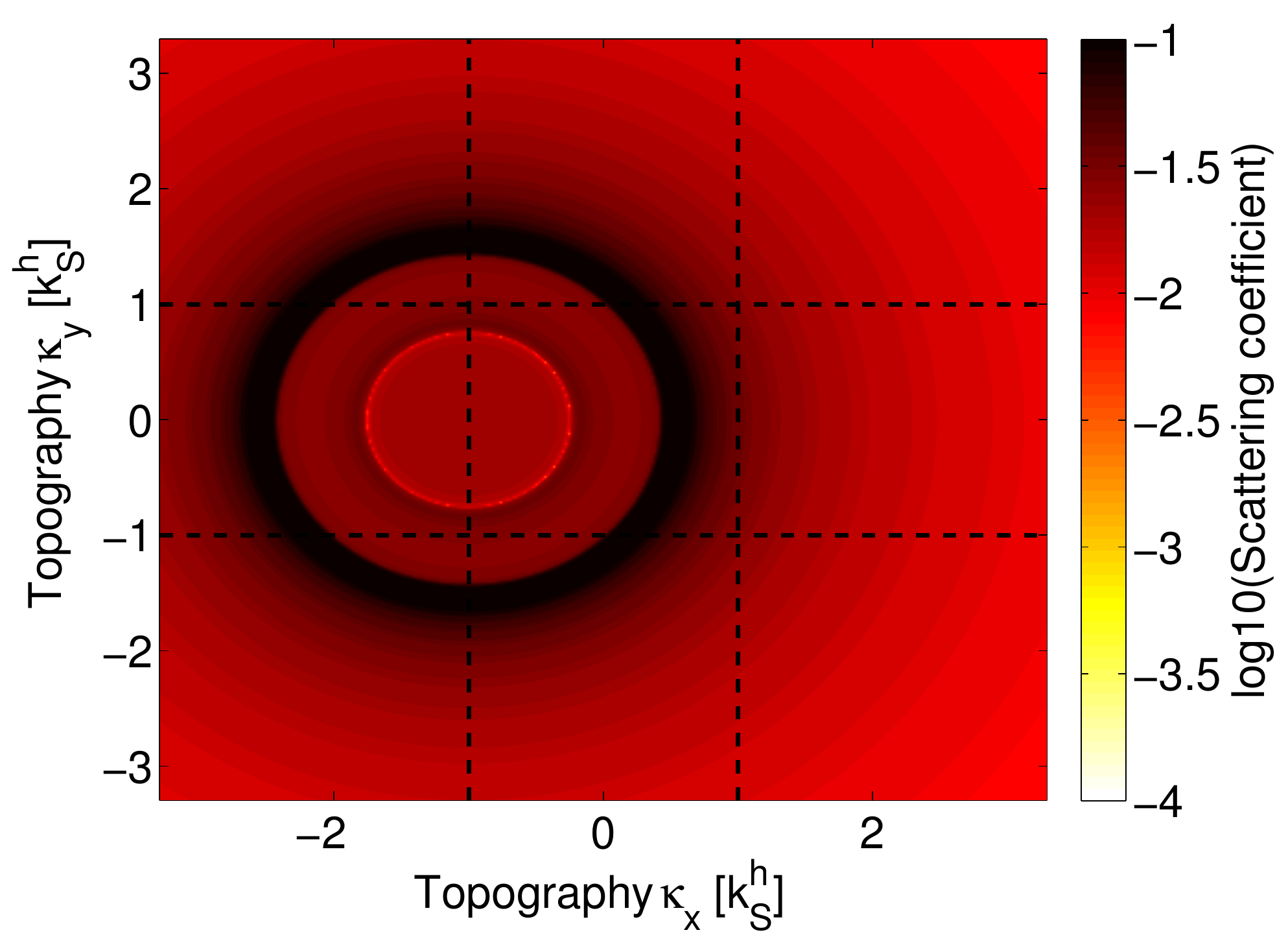}
 \caption{The plots show absolute values of the scattering coefficients into vertical surface displacement for incident compressional (left) and shear (right) waves propagating along the x-direction as a function of wave vector $\vec \kappa$ of a sinusoidal surface. The topographic wave vector is expressed in units of the horizontal wavenumber of the incident compressional or shear wave. The incident shear wave is in so-called SV polarization, which has no displacement component along the horizontal direction. In contrast to Rayleigh-wave scattering as shown in figure \ref{fig:scattR}, bulk-wave scattering coefficients also exhibit forward scattering into the Rayleigh poles that lie on the dark circle. }
 \label{fig:scattPS}
\end{figure*}

The amplitudes of the scattered field at the origin $\vec x=\vec 0$ are plotted in figures \ref{fig:scattR} and \ref{fig:scattPS}. Figure \ref{fig:scattR} also shows an example of a scattered displacement field with incident Rayleigh wave. In all cases, the amplitudes are calculated for incident waves with 1\,m amplitudes (1\,m vertical displacement for incident Rayleigh waves) and 1\,m amplitude of the sinusoidal topography. The compressional-wave to shear-wave speed ratio is slightly smaller than 2 for all plots in this section.
Certain surface wavelengths $\lambda_{\rm s}$ produce scattered waves with horizontal wavenumbers $k^{\rm h}_{\rm R}$ that correspond to the (real-valued) zeros of the Rayleigh function 
\beq
R(k^{\rm h})=((k^{\rm h})^2 - (k^{\rm v}(\beta))^2)^2 + 4 (k^{\rm h})^2 k^{\rm v}(\alpha)k^{\rm v}(\beta)
\label{eq:Rayleigh}
\eeq
The Rayleigh function appears in the denominator in coupling coefficients between seismic sources and fields near surfaces and so it also naturally emerges when solving the surface boundary equations (\ref{eq:stresscompr}) and (\ref{eq:stressshear}) for the displacement fields. 
In general, the zeros of the Rayleigh function (usually called Rayleigh poles since $R(k)$ is mostly found in denominators) describe a resonance between compressional and shear waves near the surface. Near these points, the scattering coefficients go to infinity, and the Born approximation breaks down. This problem is partially a result of considering scattering from infinite sinusoidal topographies, which is certainly a rather artificial scenario. Often, the Rayleigh poles do not yield pathological behavior when integrating over a range of wavenumbers (see for example \cite{AkRi2009}). For the purpose of this paper, we will simply understand scattering into the Rayleigh poles as very efficient without further quantifying the scattering coefficients. Finally, we want to point out that the absolute value of the scattering coefficient is also a function of frequency. The plots in figures \ref{fig:scattR} and \ref{fig:scattPS} were evaluated at $f=10\,$Hz. There is no simple frequency scaling valid for the entire wavenumber space, but scattering always becomes weaker at smaller frequencies if all other parameters are kept constant.

\section{Topographic data and site selection}
\label{sec:topodata}

As noted, we use information about surface topography to investigate seismic scattering at two representative sites. As the arms of current GW detectors are on the order of kilometers, we will analyze locations of 10km x 10km area. We have acquired surface topography data covering the United States from the USGS National Elevation Dataset. The elevation data is provided in blocks of 1$^\circ$ latitude by 1$^\circ$ longitude with 1 second resolution. This high resolution data is necessary to understand how rapidly the elevation changes over the course of land data on the scales of which we are concerned in the context of topographic scattering.

\begin{figure}[t]
 \centering 
 \includegraphics[width=5in]{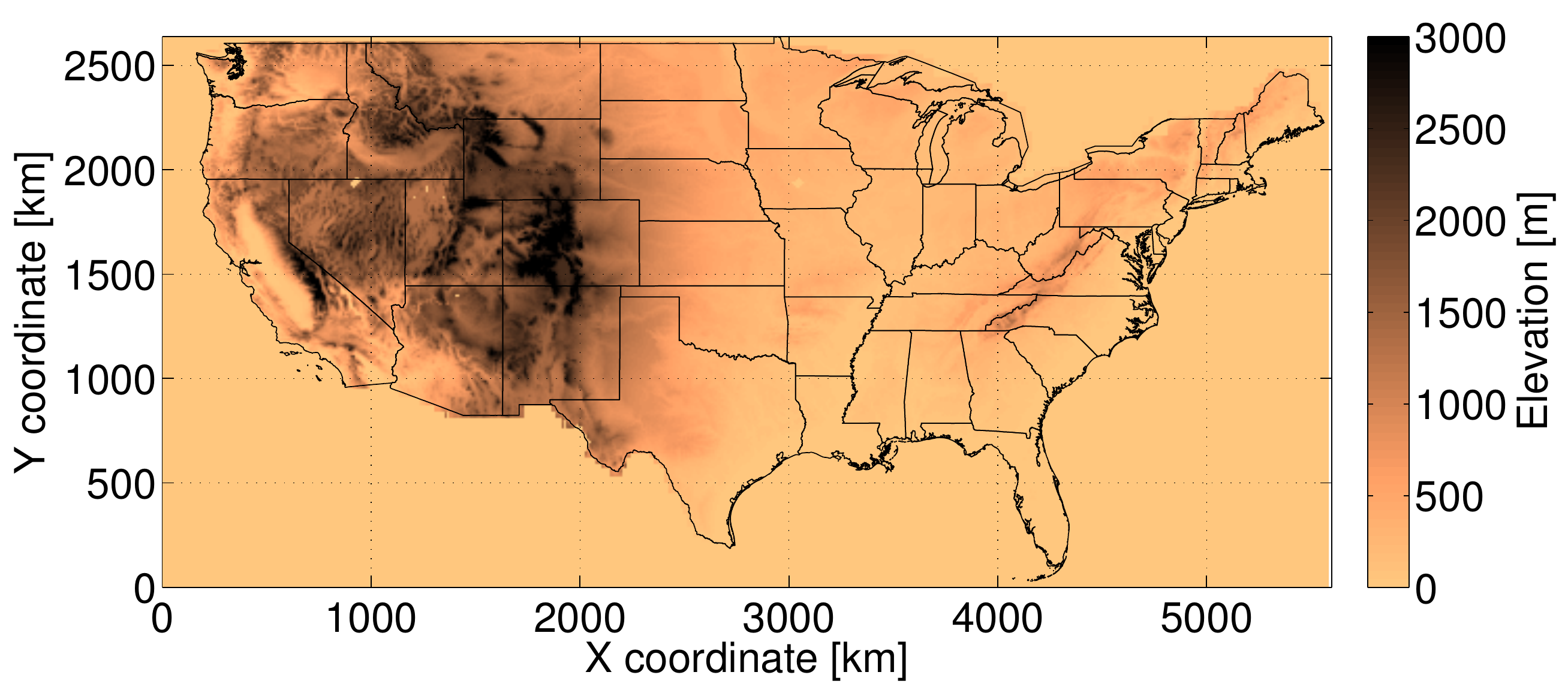}\\
 \includegraphics[width=5in]{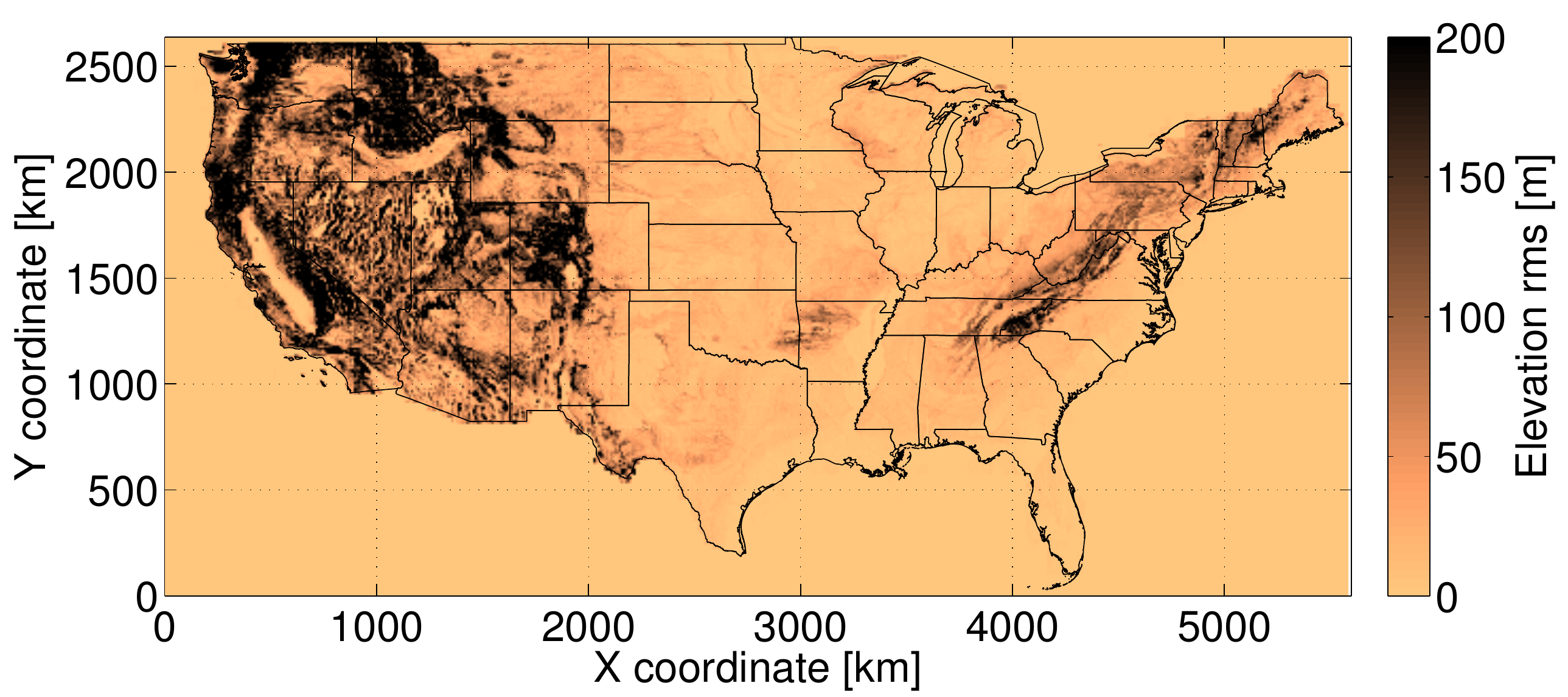}
 \caption{The plot on the top shows the elevation map of the US. Although not further studied in this paper, one important observation is that seismic noise especially above 1\,Hz is weaker in regions with high elevation. In terms of site selection, the challenge is to find a high-elevation site (i.~e.~low seismic noise) with small changes in elevation over an area of about 10\,km$\times$10\,km to minimize scattering. For this purpose, the bottom map shows the rms of the elevation calculated for areas of this size. Every point of the plot represents one of these squares so any light colors in the rms plot within high-elevation areas is a possible candidate for a low-seismic noise, low scattering site.}
 \label{fig:Topography}
\end{figure} 

As blocks of latitude and longitude are neither regularly spaced nor equal area, it is necessary to perform a coordinate transition to analyze blocks of the same area. Therefore, to map the elevation data, we use Lambert's equal-area cylindrical projection. After projection, the data is interpolated to be on a regular 10km x 10km grid covering the country. The benefits of this grid are that the arms of current GW detectors are on the order of kilometers, and so it is fair to think of each block in the grid as a location for a future GW detector. We provide two plots in  figure \ref{fig:Topography}. The top one is the mean elevation at each grid point. The bottom plot shows the root-mean-square (rms) of the elevation at each grid point. This map will be used to identify locations of low rms. This is important as locations with rapidly changing elevation will show stronger scattering. Therefore, we desire to pick locations low in elevation rms.

\begin{figure}[t]
 \centering \subfigure{}
 {\footnotesize F13A (high rms)\hspace{5cm} K07A (low rms)}\\
 \includegraphics[width=3in]{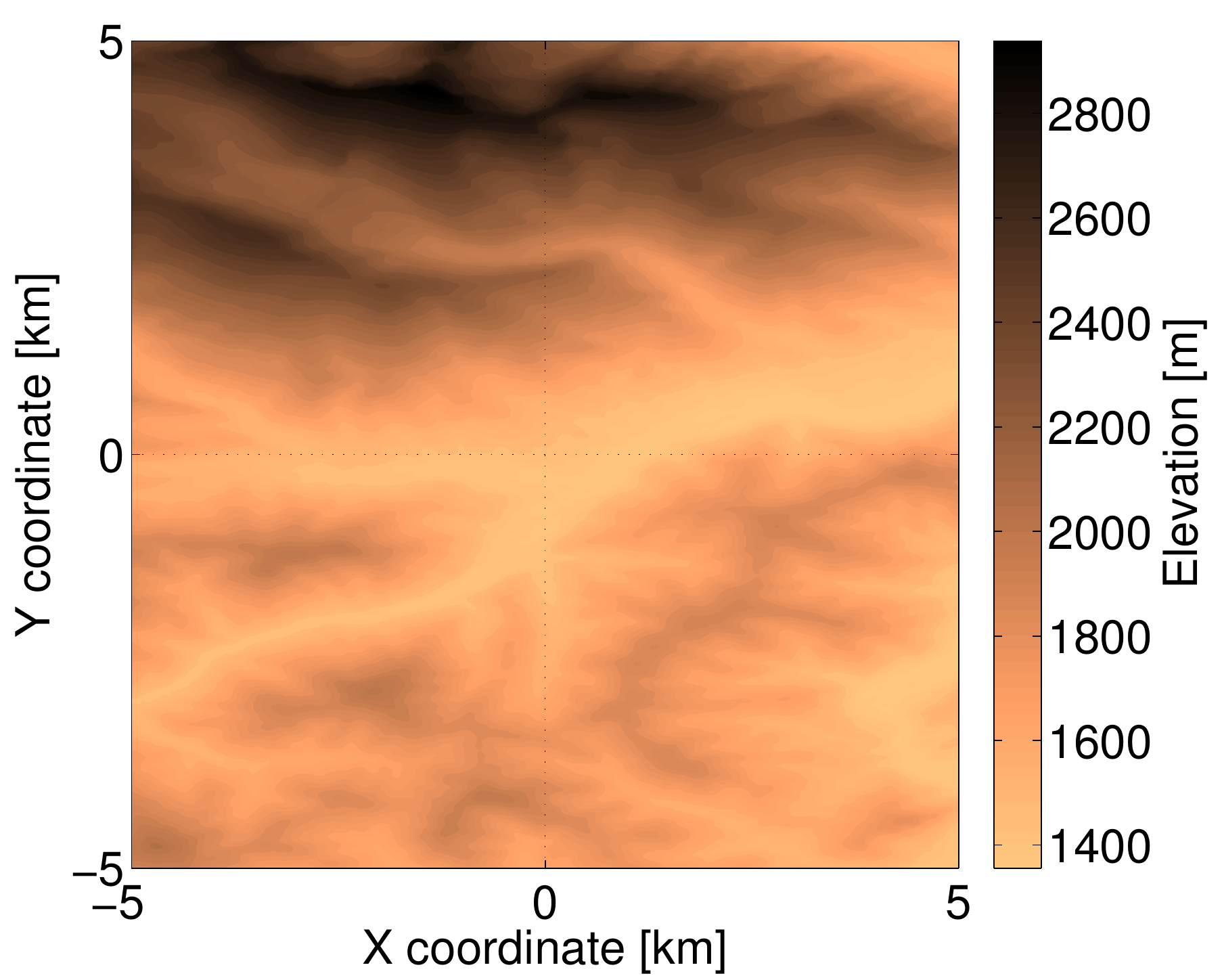}
 \centering \subfigure{}
 \includegraphics[width=3in]{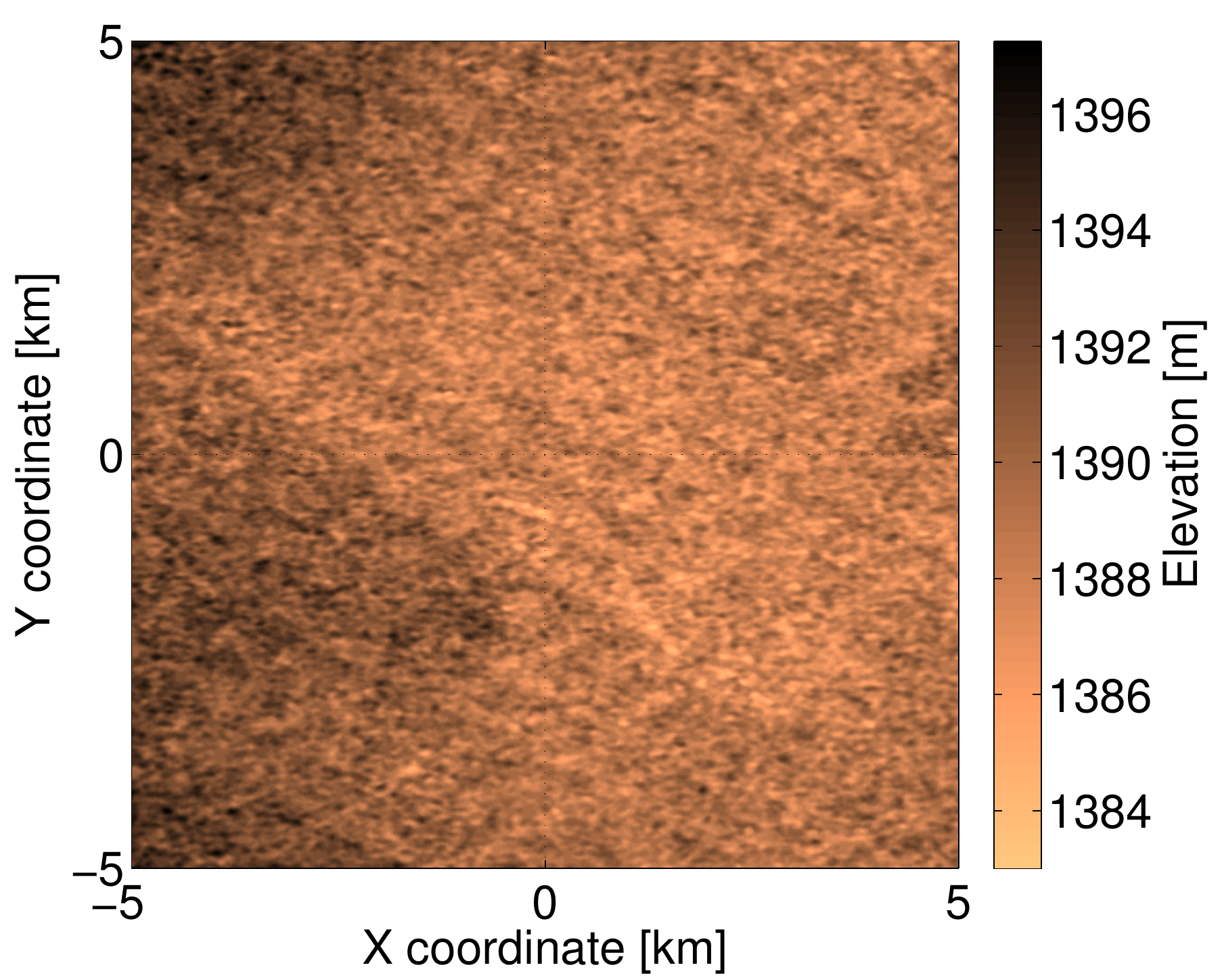}
 \caption{The two maps display elevation data of low-rms and high-rms locations. A low-rms location with weak seismic noise would be an ideal site candidate for a future GW detector. Here, we only study the scattering problem, but both sites also have known seismic noise measured by the USArray stations F13A for the map to the left, and K07A for the map to the right. Both sites have weak seismic noise at frequencies between 10\,mHz and 10\,Hz. }
 \label{fig:TopoSites}
\end{figure}

\begin{figure}[ht!]
 \centering 
 \includegraphics[width=5in]{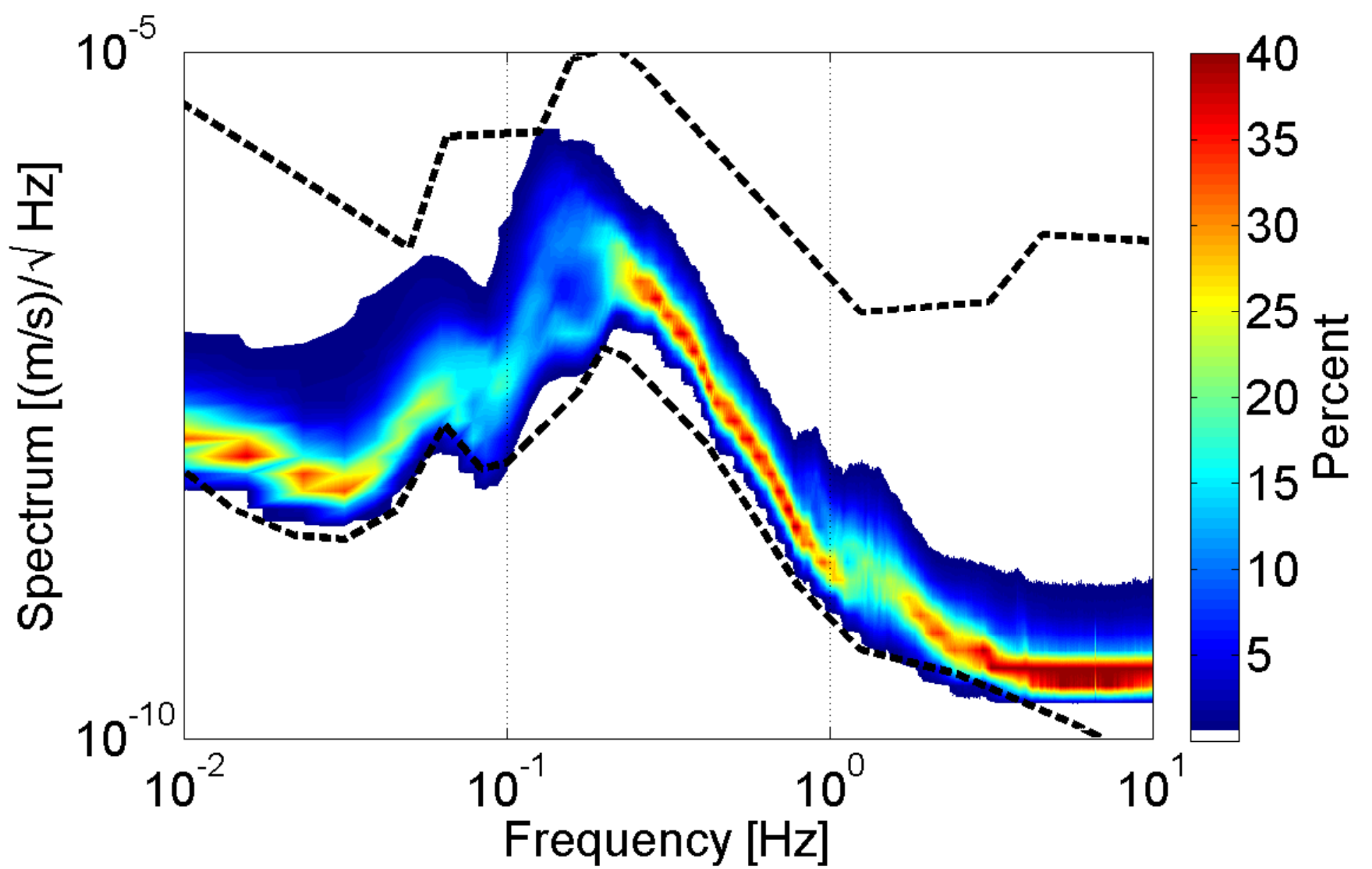}
 \caption{The plot shows the seismic spectral histogram measured at station K07A of the transferrable array (part of the USArray) based on two years of 3-hour averaged spectra. The two dashed lines are the high-noise and low-noise global seismic models. A more careful seismic analysis needs to be carried out to fully characterize a site, but the spectrum proves that flat surface, low-noise sites exist in the US.}
 \label{fig:SpecK07A}
\end{figure}

However, topography is not the only determinant of the complexity of ambient seismic fields, and this must also be factored when choosing sites. In general, flat land is better populated than land that has rapidly changing elevation. Therefore, land outside of the mountains has the shortcoming of increased anthropogenic seismic noise. For this reason, we desire a location that has low-rms and is therefore flat, while also seismically quiet. We use the rms map as a first figure-of-merit to identify promising sites, and these locations will be examined more carefully with detailed scattering calculations in the next section. Their elevation maps are shown in figure \ref{fig:TopoSites}. One site represents a high-rms area, the other one a low-rms area. The maps are centered within a few km to seismic stations of the USArray to ensure that we have local seismic data. For example, the low-rms site is near the station K07A of the transferrable array (TA), where noise spectra close to the low-noise model are observed (see \cite{Pet1993} for the definition of the seismic-noise models). The center of its map shown in figure \ref{fig:TopoSites} is located at N42.64, W119.13 in Oregon close to the border with California.
The location of the station F13A coincides with the center of the high-rms site at N45.79, W114.33 in Montana. It also represents a low-noise site. As can be seen from the map, K07A has a peak-to-peak elevation change of about 14\,m over 5\,km distance to the center in any direction, which is much less than we had initially anticipated for any high-elevation site. Together with the seismic spectrum, as shown in figure \ref{fig:SpecK07A}, it is demonstrated that low-noise, flat surface sites exist in the US that could be considered as locations for a future GW detector. We want to emphasize though that there are many more sites that can be found with similar characteristics. The Oregon site was just the best one according to our simplified rms criterion. In addition, it is certainly true that topographic scattering and seismic noise are only two of a large number of site-selection criteria. 

\section{Phenomenological scattering analysis for F13A and K07A}
\label{sec:scattsite}

The last step is to expand the topographic maps of the two selected sites into sums over sinusoidal surface perturbations and to calculate the amplitudes of scattered waves. Because the mode composition of the seismic fields at the sites is unknown, we assume that the unperturbed seismic field is composed of fundamental Rayleigh waves. This simplified model is sufficient to discuss how topographic scattering affects Newtonian-noise estimation and subtraction in future GW detectors.

\begin{figure}[t]
 \centering \subfigure{}
 {\footnotesize F13A (high rms)\hspace{5cm} K07A (low rms)}\\
 \includegraphics[width=3in]{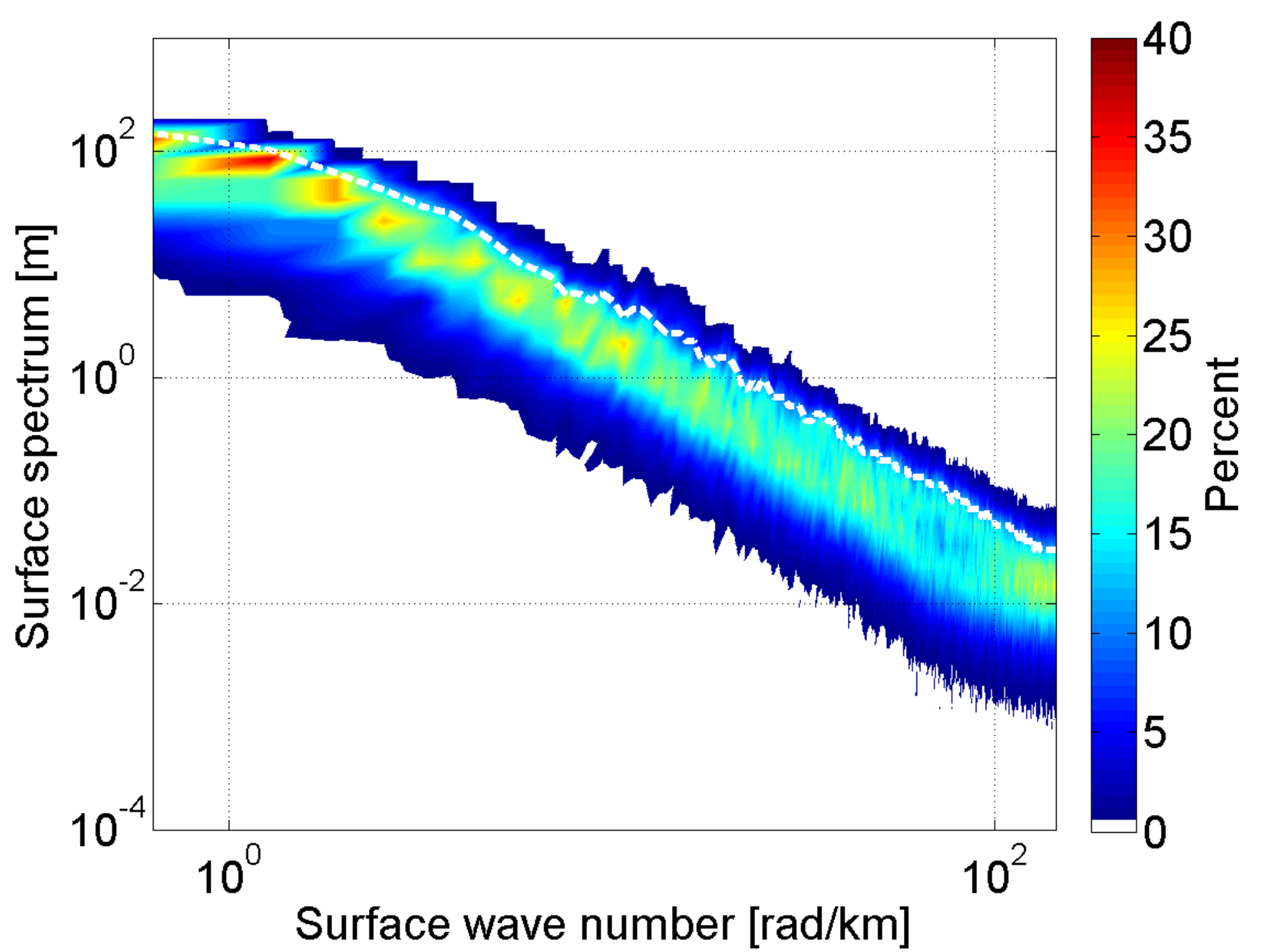}
 \centering \subfigure{}
 \includegraphics[width=3in]{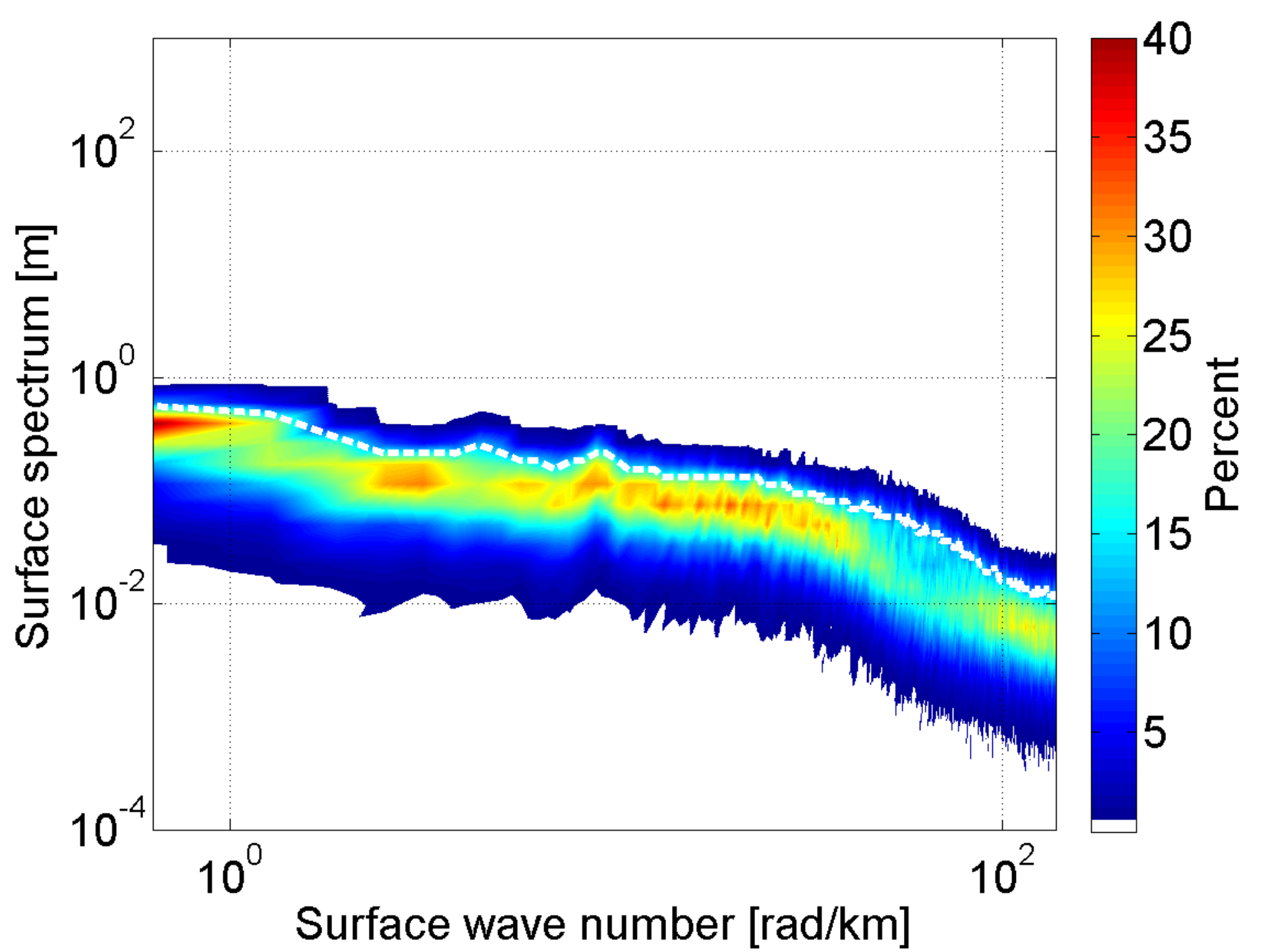}
 \caption{The spectral histograms are derived from 1D Fourier transforms of the topographic maps (figure \ref{fig:TopoSites}) along all grid lines in two orthogonal horizontal directions. The white dashed curves are the 90th percentiles of the histogram.}
 \label{fig:SpecHisto}
\end{figure}

Figure \ref{fig:SpecHisto} shows the histograms of 1D spectra of the topographic maps along all grid lines in two orthogonal directions. As scattering coefficients in wavenumber space are proportional to the topographic spectrum, the widths of the spectral histograms give an idea of how much variation in scattering is to be expected as a function of propagation direction of the incident seismic wave. The units of the spectra are chosen such that a simple sum over the harmonic components yield the 1D topography along grid lines. In this way, we can test easily if the conditions for the Born approximation still hold. For example, scattering of a 10\,m long Rayleigh wave from 1\,km long topographic features at site F13A cannot be described in the Born approximation since the corresponding topographic amplitude is larger than the length of the incident wave. Comparing the two histograms we find that the amplitudes at larger wavenumbers have similar values whereas the amplitudes at small wavenumbers differ by almost two orders of magnitude. In general, we should always expect topographic spectra to decrease with increasing $\kappa$ (the higher the mountain the wider its base). Therefore, small-$\kappa$ scattering should be more pronounced.

\begin{figure}[t]
 \centering \subfigure{}
 {\footnotesize F13A (high rms)\hspace{5cm} K07A (low rms)}\\
 \includegraphics[width=3in]{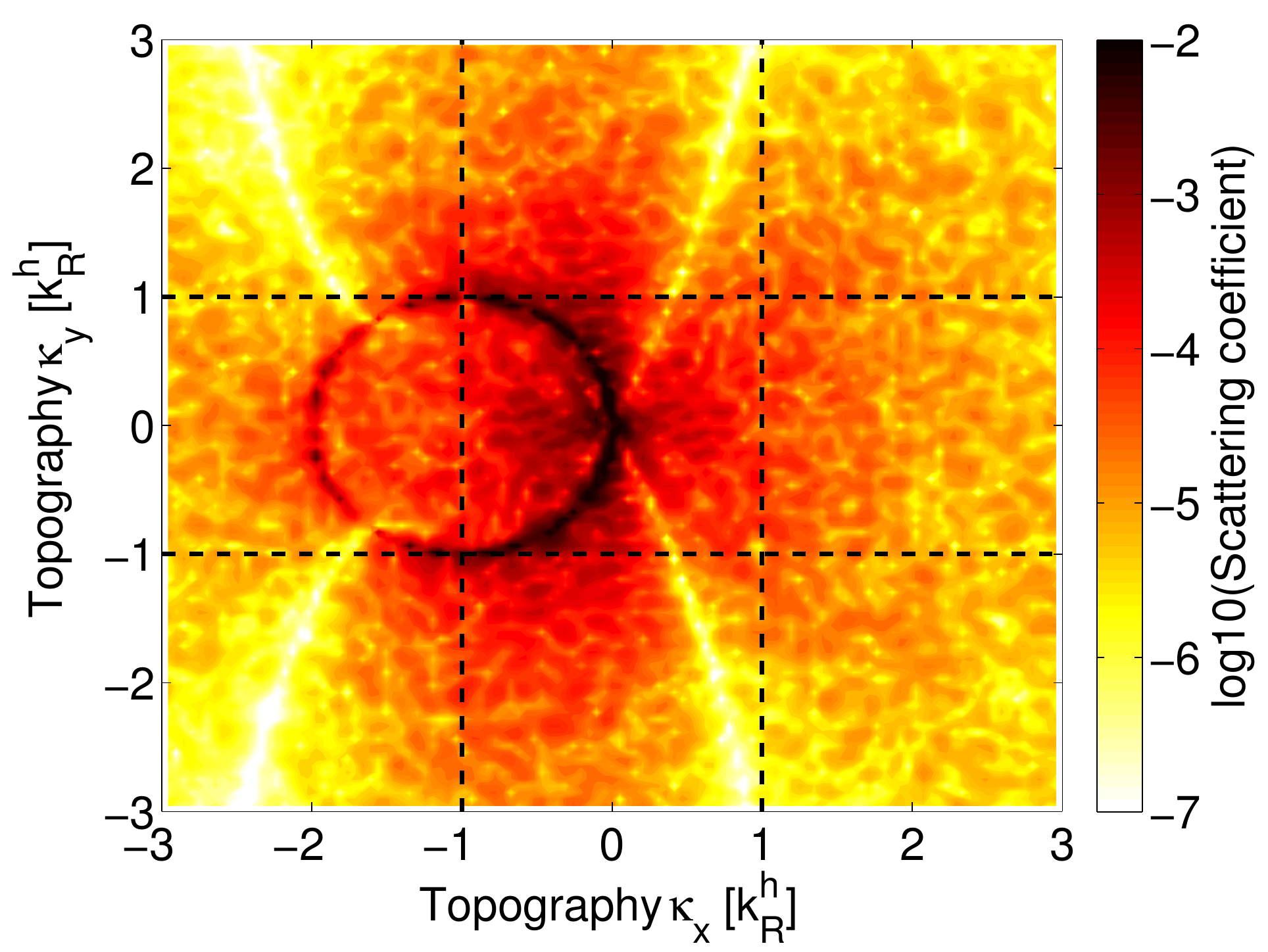}
 \centering \subfigure{}
 \includegraphics[width=3in]{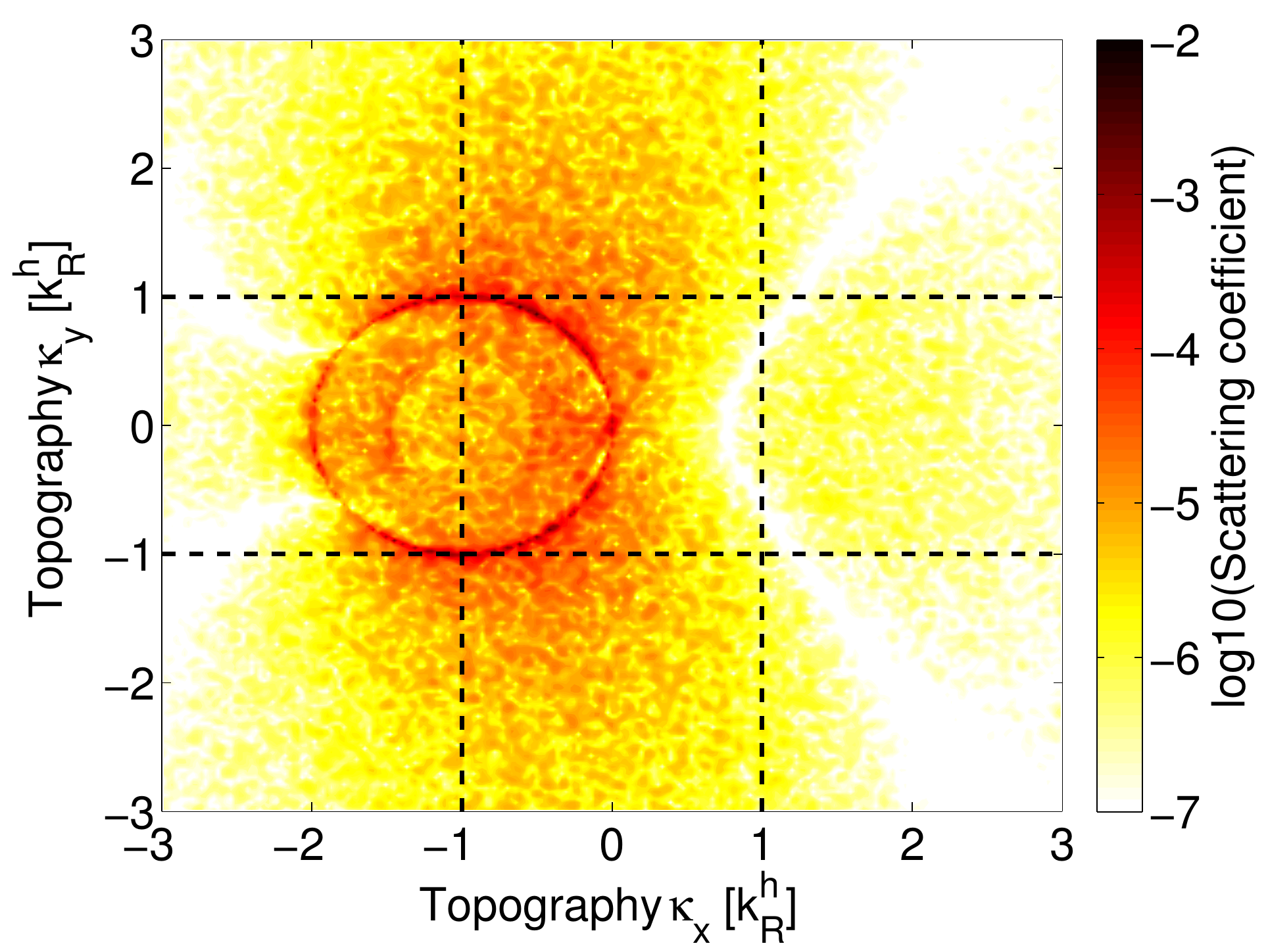}
 \caption{The contour plots show the scattering coefficients for incident Rayleigh waves. Very small surface wavenumbers cause a weak broadening of seismic spatial spectra. Scattering from very high surface wavenumbers can potentially have large effects on spatial spectra of seismic fields, but the corresponding scattered waves tend to have negligible amplitudes as can be seen in the two plots (even at the high-rms site). The wavenumber range between these two extremes is most interesting, and it also contains the poles of the scattering coefficients in the case of incident Rayleigh waves. Even though a total integrated scatter cannot be obtained due to the Rayleigh poles, scatter coefficients can be summed if one excludes a ring that lies on the poles. The results obtained for a ring of width d$k=0.01\,k_{\rm R}$ are 0.002 for the low-rms site and 0.04 for the high-rms site.}
 \label{fig:Scatt-kf}
\end{figure}

Next, we multiply the scattering coefficient shown in figure \ref{fig:scattR} with the 2D topographic spectra of each site. The result is the amplitude of scattered waves as a function of topographic wavenumber $\kappa$ normalized by the amplitude of the incident Rayleigh wave. For this analysis, we chose to calculate the scattering at lower frequency, i.~e.~at 3\,Hz, and to be more careful about the estimate of compressional and shear-wave speeds $\alpha,\,\beta$ at the two sites (which also determine the Rayleigh-wave speed). According to USGS geologic maps (http://tin.er.usgs.gov/geology/state/), the geologic unit at the low-rms Oregon site K07A is ''unconsolidated to semi-consolidated clay, silt, sand or gravel''. Speed of compressional waves in these materials is typically $\alpha=1800\,$m/s with comparatively low shear-wave speeds. We assume $\beta=900\,$m/s, which falls into the shear-wave range specified in \cite{Ore1998} for stiff soil. F13A lies in the Idaho Batholith, which contains ``faintly gneissic quartz monzonite, granodiorite, and similar rocks''. Depending on the state of weathering of the rock, compressional wave speed can range from 2000\,m/s to 5000\,m/s. Because the area also contains alluvial deposits, we chose $\alpha=2500\,$m/s near the lower end of the range and shear-wave speed $\beta=1800\,$m/s. The scattering coefficients are shown in figure \ref{fig:Scatt-kf}. As expected, scattering is significantly stronger at F13A. As the (horizontal) wave vector of a scattered wave is simply the sum of the wave vector of the incident Rayleigh wave propagating along the x-direction and the topographic wave vector $\vec\kappa$, the plots can be understood as the spatial spectra of a seismic wave that entered the site from the west as a plane wave and has just left it in the east. As we mentioned before, scattering of Rayleigh waves into Rayleigh waves has to be much weaker than suggested by our results since the Born approximation breaks down near the Rayleigh poles giving rise to infinite scattering coefficients. Neglecting scattering at these wave vectors, the main effect of irregular surface topography in the Born approximation is to broaden narrow features in the spatial spectrum of the incident wave. Due to the Rayleigh poles, this spreading should occur into arc like shapes.

We want to point out that scattering into Rayleigh waves does not influence the performance of a NN subtraction filter. The seismic array is already designed to allow subtraction of NN from Rayleigh waves. It does not make any difference whether these waves are incident from outside or produced by scattering. However, broadening of narrow features in wavenumber spectra can potentially limit NN subtraction because it would be challenging to design a seismic array that provides data to correctly estimate seismic amplitudes over continuous intervals of wavenumbers (at each frequency). Even though a total integrated scatter cannot be obtained due to the Rayleigh poles, scatter coefficients can be summed if one excludes a ring that lies on the poles. Excluding a ring of width d$k=0.01\,k_{\rm R}$, we find a total integrated scatter of 0.04 for the F13A site. Similarly, the total integrated scatter at the low-rms site K07A is 0.002. If the goal is to reduce NN by a factor 100, then scattering can not be neglected at the F13A site.

\section{Conclusion}
\label{sec:Conclude}

We discussed topographic scattering in the context of GW-detector site selection and Newtonian-noise subtraction. We found that the total contribution of waves scattered from topography can exceed values of 0.01, which makes topographic scattering relevant to NN subtraction in future low-frequency GW detectors assuming NN reduction by a factor 100. Although scattering into Rayleigh waves cannot be described in the Born approximation, we explained that this relatively efficient scattering channel would not pose a problem for NN subtraction independent of the value of the scattering coefficient. Since it is possible that scattering into Rayleigh waves significantly dissipates energy from the incident wave at strong scattering sites, the absolute values of scattering coefficients may be overestimated though. This issue can for example be resolved with numerical simulations of scattering into Rayleigh waves. For scattering into wavenumbers that do not correspond to Rayleigh poles, the outlined theory provides robust predictions of scattering coefficients. Since scattering into body fields is negligible for all types of incident waves, the problem of topographic scattering can be avoided by constructing the detector sufficiently deep underground. This scenario would have to be studied carefully though since scattered surface waves are not generally of Rayleigh type and can therefore generate displacement deeper than known for Rayleigh fields.

The main challenge with subtraction of NN from surface waves is the design of the seismic array. The requirement for the array diameter is derived from the longest seismic wavelengths in the field at lowest frequency, whereas the array density, which will be variable throughout the array, depends on the shortest wavelengths at highest frequency. In fields where there is no unique correspondence between frequency and wavelength, it will be very difficult to design a seismic array for NN subtraction. As was shown, this situation occurs when a site exhibits strong topographic scattering that causes amplitudes in wavenumber space to spread over adjacent wave vectors. This spreading happens with relatively high scattering amplitudes because it is associated with the lowest wavenumber components of the topography that typically have the largest amplitudes. We have not repeated the same analysis for scattering of bulk waves since the effects are very similar. In addition, scattering of bulk waves into Rayleigh poles is generally driven by higher-wavenumber components of the surface topography that have weaker amplitudes.

In the future, a simulation of topographic scattering should be carried out that includes the seismometer array and the noise subtraction. Then one can address questions such as how the array needs to be adjusted to minimize the effect of small-wavelength scattering on subtraction residuals. Although we found that topographic scattering can be negligible at flat sites or that it does not lead to adverse effects, it is certainly desirable to understand how to design arrays to achieve subtraction goals at arbitrary sites. 

\section{Acknowledgments}
The authors gratefully acknowledge the support of the United States National Science Foundation for the construction and operation of the LIGO Laboratory. M.~C.'s work was funded by the NSF through the California Institute of Technology's Summer Undergraduate Fund. We thank Jennifer Driggers, Vuk Mandic and Gary Pavlis for many helpful discussions about seismic scattering.\\[1cm]

\raggedright
\bibliography{references}

\begin{thebibliography}{21}
\expandafter\ifx\csname natexlab\endcsname\relax\def\natexlab#1{#1}\fi
\expandafter\ifx\csname bibnamefont\endcsname\relax
  \def\bibnamefont#1{#1}\fi
\expandafter\ifx\csname bibfnamefont\endcsname\relax
  \def\bibfnamefont#1{#1}\fi
\expandafter\ifx\csname citenamefont\endcsname\relax
  \def\citenamefont#1{#1}\fi
\expandafter\ifx\csname url\endcsname\relax
  \def\url#1{\texttt{#1}}\fi
\expandafter\ifx\csname urlprefix\endcsname\relax\def\urlprefix{URL }\fi
\providecommand{\bibinfo}[2]{#2}
\providecommand{\eprint}[2][]{\url{#2}}

\bibitem[{\citenamefont{{LIGO Scientific Collaboration}}(2011)}]{aLIG2011}
\bibinfo{author}{\bibnamefont{{LIGO Scientific Collaboration}}},
  \bibinfo{journal}{{LIGO M060056-v2}}  (\bibinfo{year}{2011}).

\bibitem[{\citenamefont{{The Virgo Collaboration}}(2009)}]{aVir2009}
\bibinfo{author}{\bibnamefont{{The Virgo Collaboration}}},
  \bibinfo{journal}{{VIR 027A 09}}  (\bibinfo{year}{2009}).

\bibitem[{\citenamefont{(on behalf of~the
  LCGT~Collaboration)}(2010)}]{KuEA2010}
\bibinfo{author}{\bibfnamefont{K.~K.} \bibnamefont{(on behalf of~the
  LCGT~Collaboration)}}, \bibinfo{journal}{Class.~Quantum Grav.}
  \textbf{\bibinfo{volume}{27}}, \bibinfo{pages}{084004}
  (\bibinfo{year}{2010}).

\bibitem[{\citenamefont{{S.~Hild et al.}}(2011)}]{HiEA2011}
\bibinfo{author}{\bibnamefont{{S.~Hild et al.}}},
  \bibinfo{journal}{Class.~Quantum Grav.} \textbf{\bibinfo{volume}{28}},
  \bibinfo{pages}{094013} (\bibinfo{year}{2011}).

\bibitem[{\citenamefont{Team}(2011)}]{ET2011}
\bibinfo{author}{\bibfnamefont{E.~S.} \bibnamefont{Team}},
  \textbf{\bibinfo{volume}{ET-0106C-10}} (\bibinfo{year}{2011}).

\bibitem[{\citenamefont{Hughes and Thorne}(1998)}]{HuTh1998}
\bibinfo{author}{\bibfnamefont{S.~A.} \bibnamefont{Hughes}} \bibnamefont{and}
  \bibinfo{author}{\bibfnamefont{K.~S.} \bibnamefont{Thorne}},
  \bibinfo{journal}{Phys.~Rev.~D} \textbf{\bibinfo{volume}{58}},
  \bibinfo{pages}{122002} (\bibinfo{year}{1998}).

\bibitem[{\citenamefont{{M.~Beccaria et al}}(1998)}]{BeEA1998}
\bibinfo{author}{\bibnamefont{{M.~Beccaria et al}}},
  \bibinfo{journal}{Class.~Quantum Grav.} \textbf{\bibinfo{volume}{15}},
  \bibinfo{pages}{3339} (\bibinfo{year}{1998}).

\bibitem[{\citenamefont{Creighton}(2008)}]{Cre2008}
\bibinfo{author}{\bibfnamefont{T.}~\bibnamefont{Creighton}},
  \bibinfo{journal}{Class.~Quantum Grav.} \textbf{\bibinfo{volume}{25}},
  \bibinfo{pages}{125011} (\bibinfo{year}{2008}).

\bibitem[{\citenamefont{Thorne and Winstein}(1999)}]{ThWi1999}
\bibinfo{author}{\bibfnamefont{K.~S.} \bibnamefont{Thorne}} \bibnamefont{and}
  \bibinfo{author}{\bibfnamefont{C.~J.} \bibnamefont{Winstein}},
  \bibinfo{journal}{Phys.~Rev.~D} \textbf{\bibinfo{volume}{60}},
  \bibinfo{pages}{082001} (\bibinfo{year}{1999}).

\bibitem[{\citenamefont{{M.~G.~Beker et al.}}(2010)}]{BeEA2010}
\bibinfo{author}{\bibnamefont{{M.~G.~Beker et al.}}}, \bibinfo{journal}{Gen
  Relativ Gravit}  (\bibinfo{year}{2010}).

\bibitem[{\citenamefont{{J.~Harms et al}}(2010)}]{HaEA2010}
\bibinfo{author}{\bibnamefont{{J.~Harms et al}}},
  \bibinfo{journal}{Class.~Quantum Grav.} \textbf{\bibinfo{volume}{27}},
  \bibinfo{pages}{225011} (\bibinfo{year}{2010}).

\bibitem[{\citenamefont{Harms et~al.}(2009{\natexlab{a}})\citenamefont{Harms,
  DeSalvo, Dorsher, and Mandic}}]{HaEA2009a}
\bibinfo{author}{\bibfnamefont{J.}~\bibnamefont{Harms}},
  \bibinfo{author}{\bibfnamefont{R.}~\bibnamefont{DeSalvo}},
  \bibinfo{author}{\bibfnamefont{S.}~\bibnamefont{Dorsher}}, \bibnamefont{and}
  \bibinfo{author}{\bibfnamefont{V.}~\bibnamefont{Mandic}},
  \bibinfo{journal}{Phys.~Rev.~D} \textbf{\bibinfo{volume}{80}},
  \bibinfo{pages}{122001} (\bibinfo{year}{2009}{\natexlab{a}}).

\bibitem[{\citenamefont{Harms et~al.}(2009{\natexlab{b}})\citenamefont{Harms,
  DeSalvo, Dorsher, and Mandic}}]{HaEA2009b}
\bibinfo{author}{\bibfnamefont{J.}~\bibnamefont{Harms}},
  \bibinfo{author}{\bibfnamefont{R.}~\bibnamefont{DeSalvo}},
  \bibinfo{author}{\bibfnamefont{S.}~\bibnamefont{Dorsher}}, \bibnamefont{and}
  \bibinfo{author}{\bibfnamefont{V.}~\bibnamefont{Mandic}},
  \bibinfo{journal}{arXiv:0910.2774}  (\bibinfo{year}{2009}{\natexlab{b}}).

\bibitem[{\citenamefont{Harms and O'Reilly}(2011)}]{HaOR2011}
\bibinfo{author}{\bibfnamefont{J.}~\bibnamefont{Harms}} \bibnamefont{and}
  \bibinfo{author}{\bibfnamefont{B.}~\bibnamefont{O'Reilly}},
  \bibinfo{journal}{BSSA} \textbf{\bibinfo{volume}{101}}, \bibinfo{pages}{1478}
  (\bibinfo{year}{2011}).

\bibitem[{\citenamefont{Gilbert and Knopoff}(1960)}]{GiKn1960}
\bibinfo{author}{\bibfnamefont{F.}~\bibnamefont{Gilbert}} \bibnamefont{and}
  \bibinfo{author}{\bibfnamefont{L.}~\bibnamefont{Knopoff}},
  \bibinfo{journal}{J.~Geophys.~Res.} \textbf{\bibinfo{volume}{65}},
  \bibinfo{pages}{3437} (\bibinfo{year}{1960}).

\bibitem[{\citenamefont{Abubakar}(1962)}]{Abu1962}
\bibinfo{author}{\bibfnamefont{I.}~\bibnamefont{Abubakar}},
  \bibinfo{journal}{Math.~Proc.~Cambridge Phil.~Soc.}
  \textbf{\bibinfo{volume}{58}}, \bibinfo{pages}{136} (\bibinfo{year}{1962}).

\bibitem[{\citenamefont{Hudson}(1967)}]{Hud1967}
\bibinfo{author}{\bibfnamefont{J.~A.} \bibnamefont{Hudson}},
  \bibinfo{journal}{Geophys.~J.~R.~astr.~Soc.} \textbf{\bibinfo{volume}{13}},
  \bibinfo{pages}{441} (\bibinfo{year}{1967}).

\bibitem[{\citenamefont{Ogilvy}(1987)}]{Ogi1987}
\bibinfo{author}{\bibfnamefont{J.~A.} \bibnamefont{Ogilvy}},
  \bibinfo{journal}{Rep.~Prog.~Phys.} \textbf{\bibinfo{volume}{50}},
  \bibinfo{pages}{1553} (\bibinfo{year}{1987}).

\bibitem[{\citenamefont{Aki and Richards}(2009)}]{AkRi2009}
\bibinfo{author}{\bibfnamefont{K.}~\bibnamefont{Aki}} \bibnamefont{and}
  \bibinfo{author}{\bibfnamefont{P.~G.} \bibnamefont{Richards}},
  \emph{\bibinfo{title}{{Quantitative Seismology, 2nd edition}}}
  (\bibinfo{publisher}{University Science Books}, \bibinfo{year}{2009}).

\bibitem[{\citenamefont{Peterson}(1993)}]{Pet1993}
\bibinfo{author}{\bibfnamefont{J.}~\bibnamefont{Peterson}},
  \bibinfo{journal}{Open-file report} \textbf{\bibinfo{volume}{93-322}}
  (\bibinfo{year}{1993}).

\bibitem[{Ore(1998)}]{Ore1998}
\emph{\bibinfo{title}{Oregon Geology}}, vol.~\bibinfo{volume}{60}
  (\bibinfo{publisher}{Oregon Department of Geology and Mineral Industries},
  \bibinfo{year}{1998}).

\end{thebibliography}

\end{document}